# Magnetometry using sodium fluorescence with synchronous modulation of two-photon resonant light fields

**RAGHWINDER SINGH GREWAL, MAURICIO PULIDO, GOUR PATI, RENU TRIPATHI** *

Division of Physics, Engineering, Mathematics and Computer Science, Delaware State University, Dover, DE 19901, USA

**Corresponding author: rtripathi@desu.edu*

**We report a new technique for generating magnetic resonance with synchronous modulation of two-photon resonant light fields. Magnetic resonances in fluorescence from a sodium cell are measured to demonstrate suitability of this technique for remote magnetometry. A strong magnetic resonance with its dip corresponding to the Larmor frequency is produced in the presence of a transverse magnetic field. An additional resonance at $3\Omega_L$ is observed, which can be used to determine the magnetic field orientation. We have developed a theoretical model based on the density matrix equations to verify our experimental observations. An average magnetic field sensitivity of 41 $\mathrm{pT}/\sqrt{\mathrm{Hz}}$ is measured using light duty cycles ranging from 35% to 10%. We have discussed possible changes that can be made to improve the sensitivity of this scheme further.**

***OCIS codes:*** *(270.1670) Coherent optical effects; (020.1335) Atom optics; (020.7490) Zeeman effect; (300.6280) Spectroscopy, fluorescence and luminescence*

Optically pumped atomic magnetometers are extensively studied for sensitive detection of magnetic fields in biomagnetics [1,2], fundamental science [3] and geophysical [4] applications. Synchronous pumping of atoms using modulated light increases the dynamic range in magnetic field measurement from microguass level to above the earth field [5,6]. Recently, magnetic-field measurement using fluorescence from sodium is being explored with interest and practical applications in remote magnetometry [7]. Laboratory studies in sodium cells are used as simulation platforms for studying the performance in remote earth field measurement [8,9]. Several sky experiments are conducted using back-scattered fluorescence from the mesosphere generated by resonant excitation of sodium $D_2$ line with a modulated laser beam [10-13]. In this case, the earth field is measured by remotely detecting the magnetic resonance produced at the Larmor precession frequency, $\Omega_L$ of sodium atoms in the mesosphere. The origin of this resonance is explained by the well-known Bell-Bloom scheme [14]. Currently, sensitivity reported in the sky experiments is very low (with highest sensitivity reported as 28 $\mathrm{nT}/\sqrt{\mathrm{Hz}}$ [11]) compared to the sensitivity achieved in laboratory based remote sodium cell magnetometer (i.e. 150 $\mathrm{pT}/\sqrt{\mathrm{Hz}}$ achieved at $D_1$ line [8]). The sensitivity in sky experiment is primarily limited by the poor $S/N$ of the resonance signal, and the photon shot-noise associated with the weak return fluorescence. A method using an additional repump light has been proposed to pump atoms back to the target state for increasing the amplitude of resonance [11]. However, the presence of repump also causes linewidth broadening and light shift, which compromises the sensitivity [15]. Polarization modulation with alternate circular polarizations was employed in sky experiment to increase the return fluorescence by confining atoms to strong cyclic transitions [12]. This technique did not provide any apparent improvement in the magnetometer performance. Thus, new techniques need to be explored for improving the performance in remote magnetometry experiments to sub-nanotesla level, so that planetary science studies could be performed using mesospheric magnetic fields.

In this paper, we demonstrate a new technique, suitable for performing remote magnetometry using synchronous modulation of two laser fields, which are two-photon resonant with the sodium hyperfine ground states. Two-photon resonance with continuous laser fields has been widely studied for coherent population trapping (CPT), particularly for developing miniaturized CPT based atomic clocks and magnetometers [16-18]. The CPT resonance is produced by dark superposition of magnetic sublevels in the hyperfine ground states of alkali atoms. In the present work, synchronous modulation of two-photon resonant CPT fields is used as a strategy for producing magnetic resonances in fluorescence from $D_1$ line in a sodium cell. We call this as the synchronous CPT scheme. The origin of magnetic resonances produced by the synchronous CPT scheme, is explained using Λ-systems formed in the sodium $D_1$ manifold. To study the performance of synchronous CPT scheme for remote magnetometry, we measured the resonance signal at $\Omega_L$ by applying a magnetic field perpendicular to the light propagation direction. In order to find the optimal duty cycle of light modulation for attaining high performance, amplitudes and linewidths of the resonance are measured as a function of the duty cycle, by keeping the peak intensity in CPT fields constant. We observed high amplitude in $\Omega_L$ resonance at longer duty cycles (close to 35%), which could be favorable in sky experiments for producing high return fluorescence. We show that a maximum sensitivity of 37 pT/$\sqrt{\mathrm{Hz}}$ in magnetic field measurement is achieved using the $\Omega_L$ resonance signal at 25% duty cycle. Sensitivity can be further improved by modifying the modulation technique used in our experiment. An additional resonance at $3\Omega_L$ is observed using the synchronous CPT scheme, which can be used in determining the orientation of external magnetic field in addition to its strength. This resonance has no counterpart in the Bell-Bloom scheme [8,9,14]. Theoretical results are obtained using atomic density matrix analysis to verify our experimental observations.

Figure 1 shows the schematic of the experimental arrangement. A frequency-doubled Raman fiber amplifier laser with a narrow linewidth (< 1 MHz) and a maximum output power of 2W, is used in our experiment. The laser wavelength is tuned to Na $D_1$ resonance (589.756 nm) and is monitored by utilizing saturation absorption spectroscopy (SAS) in a buffer-gas-free Na reference cell. Using SAS peaks, the laser is tuned to $F_g= 2\rightarrow F_e =2$ transition in Na $D_1$ line. For implementing the synchronous CPT scheme, the laser beam is modulated by a low-frequency (MHz) acousto-optic modulator (AOM) followed by a high-frequency (GHz) electro-optic modulator (EOM). The amplitude-modulated first-order diffracted light from AOM2 is frequency shifted by 80 MHz from the laser frequency. This frequency shift is compensated using another identical AOM (AOM1) in the beam path of the SAS setup. The EOM (QUBIG, Model: PM-Na23_1.7K3) is driven by an RF signal generator and amplifier combination to create optical sidebands at frequencies ±1.771 GHz matching the hyperfine ground state frequency separation in sodium atom. The carrier to sideband power ratio is approximately set to 2:1, measured using a scanning Fabry-Perot interferometer (Thorlab, SA210-5B). The laser beam diameter is expanded from 2 mm to 8 mm using a telescopic lens configuration (lenses L2 and L3), to prevent transit time broadening. Experiments are conducted using a Na cell containing 10 Torr neon buffer gas. The laser beam intensity is controlled using a neutral density (ND) filter, and the beam polarization is adjusted to circular ($\sigma^+$) polarization using a $\lambda/4$ plate. The cell is wrapped with twisted nichrome wires to cancel the residual magnetic field produced by applied DC heating current. The Na cell is heated to 92°C to yield a low vapor density of about $1.6\times10^9$ atoms/cm$^3$. The cell is kept inside a non-magnetic chamber for thermal insulation in order to maintain a steady temperature, which is then installed at the center of a two-layered mu-metal enclosure with a shielding factor of ~$10^2$. Residual magnetic field is further cancelled using three-axis Helmholtz coils mounted inside the mu-metal enclosure. These coils are also utilized in our experiment to apply a constant field $B_y$ along the y-direction in synchronous CPT scheme. We have performed all experimental measurements by collecting fluorescence light on a PMT, placed perpendicular to the light propagation direction. An ultra-narrow band-pass filter with bandwidth 1 nm centered around 589.45 nm is kept before the PMT to remove background light from fluorescence. Magnetic resonance signals are detected using a low-pass filter (LPF).

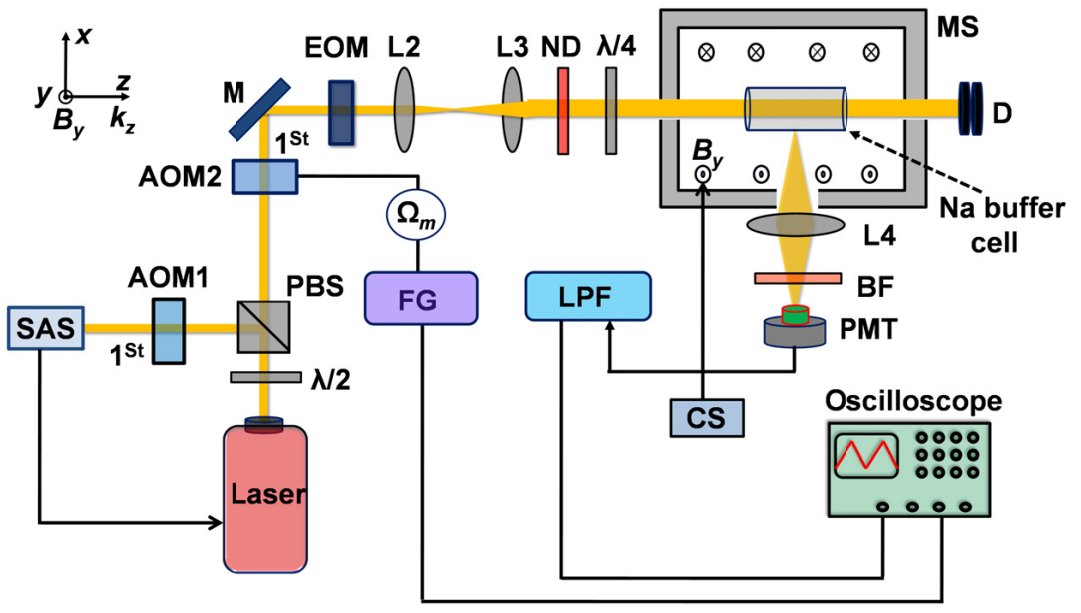

Fig. 1. Diagram showing the experimental arrangement. λ/2, half-wave plate; λ/4, quarter-wave plate; PBS, polarizer beam splitter; BS, beam splitter; D, beam dump; M, mirror; L1-L4, convex lenses; BF, band-pass filter; PMT, photomultiplier tube; FG, function generator.

Figure 2 (a-c) shows possible Λ-systems formed by two-photon resonant light fields in Na $D_1$ line in the presence of a magnetic field applied in an arbitrary direction. The Λ-systems are drawn by considering magnetic field direction as the quantization axis, in which case the light fields consist $\sigma^-$ and $\pi$ components in addition to its original $\sigma^+$polarization [19]. The two legs of Λ-system correspond to light fields generated by the EOM i.e. a carrier and a positive first-order sideband. The carrier is tuned to the $F_g$=2→$F_e$=2 transition, and the positive sideband is automatically tuned to the $F_g$=1→$F_e$=2 transition. Each Λ-system is responsible for producing a dark state corresponding to the superposition of ground state sublevels. CPT resonance occurs when frequency difference, $\nu_D$ between the carrier and the positive sideband is changed such that two-photon (or difference) detuning $\Delta\,(=\nu_D-\Delta_{hfs})$ matches with the resonant frequency of a particular Λ-system, i.e. $\Delta=0,\pm\Omega_L,\pm2\Omega_L,\pm3\Omega_L$ as shown in the diagram [19,20]. Synchronous CPT scheme is implemented by keeping the light fields (i.e. carrier and positive sideband) exactly on two-photon resonance (i.e. $\Delta=0$) and by modulating the CPT fields with frequency $\Omega_m$. In this case, different frequency components of modulated CPT fields form the same Λ-systems shown in Figures 2(a-c), and resonances will occur when $\Omega_m$ matches with $\Omega_L$, $2\Omega_L$ and $3\Omega_L$. Figure 2(d) shows a comparison with the conventional Bell-Bloom scheme, which uses a single modulated light field tuned to resonance with the $F_g$=2→$F_e$=2 transition. In this case, Λ-systems are formed by magnetic sublevels within the $F_g$=2 ground state. Resonance occurs when the modulation frequency, $\Omega_m$ matches with $\Omega_L$ and $2\Omega_L$. Compared to the synchronous CPT scheme, a $3\Omega_L$ resonance cannot be produced by the Bell-Bloom scheme. Loss of atoms to the other ground state through spontaneous emission can also reduce the contrast of resonance produced by the Bell-Bloom scheme.

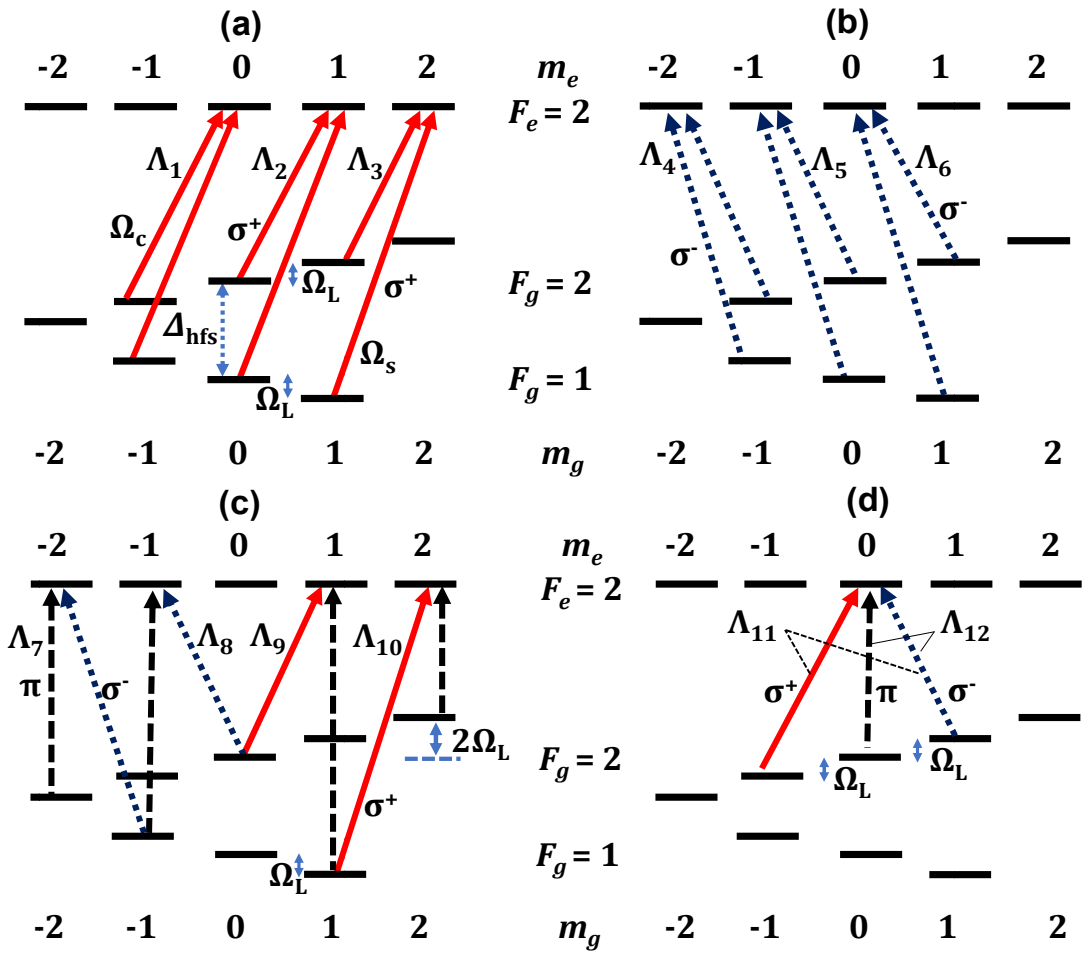

Fig. 2. Na $D_1$ line energy level diagram showing possible Λ-systems formed in the presence of a magnetic field. Λ-systems with two-photon resonant frequencies (a,b) 0 and $\pm2\Omega_L$; (c) $\pm\Omega_L$ and $\pm3\Omega_L$, and (d) Λ-systems formed in the Bell-Bloom scheme. Here $\Omega_L= \gamma B$ is the Larmor frequency, $\gamma$ is the gyromagnetic ratio. All possible Λ-systems are not shown in (c,d). The Zeeman shifts in the excited state are neglected.

Figure 3 shows the observed CPT spectrum as a function of difference detuning Δ in the presence of a fixed applied magnetic field. To improve $S/N$, the resonances are produced by deliberately modulating the laser fields with a high modulation frequency (i.e. $\Omega_m = 200\ kHz \gg \Omega_L$) and by acquiring them with low-pass filtering. The total intensity in the carrier and the sideband after the EOM is set to 5.2 W/m$^2$. In the presence of a longitudinal field ($B$=$B_z$), CPT resonances [Fig. 3 (solid red curve)] are created due to $\Lambda_1$-$\Lambda_3$ systems [shown in Fig. 2(a)], with centers of dips satisfying conditions Δ=0 and ±$2\Omega_L$. In addition to $B_z$, when a transverse magnetic field ($B_x$ and $B_y$) is applied, additional CPT resonances [Fig. 3 (dashed blue curve)] at Δ=±$\Omega_L$ and ±$3\Omega_L$ are created due to new $\Lambda_7$-$\Lambda_{10}$ systems formed by ($\sigma^\pm$, π) excitations [shown in Fig. 2(b & c)]. This increases the total number of CPT dips to seven in the spectrum. The contrast of a CPT dip depends on the population distribution among the ground state Zeeman sublevels and the coupling strengths of $\sigma^+$, $\sigma^-$ and π-transitions involved in the corresponding Λ-system [19].

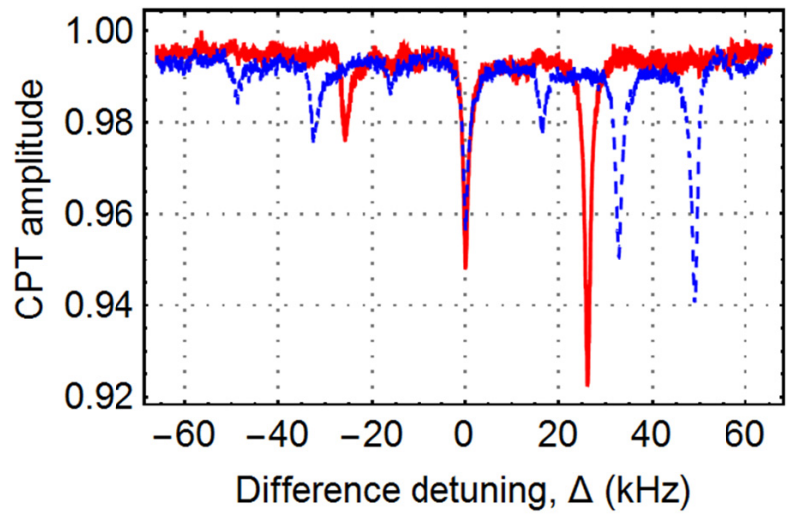

Fig. 3. CPT spectrum plotted as a function of difference detuning Δ between the CPT fields for $B_z$= 18 mG, $B_x$= $B_y$=0 (solid red line) and $B_z$= 18 mG, $B_x$= 8.5 mG, $B_y$= 8.8 mG (dashed blue line).

Next, we measure magnetic resonances produced using the synchronous CPT scheme. In this experiment, the CPT fields are kept on two-photon resonance (i.e. Δ=0) and modulated with frequency $\Omega_m$ corresponding to the Larmor frequency $\Omega_L$. Resonances are acquired by scanning $\Omega_m$ over a wide range (40 kHz to 200 kHz), for a fixed higher applied transverse magnetic field $B_y$ = 85.7 mG ($\Omega_L$=60 kHz) [Fig. 4a]. At 50% duty cycle, the modulated light has dominant first-harmonics, which form Λ-systems [shown in Fig. 2(c)] along with the unmodulated CPT fields. Figure 4a (blue curve) shows strong magnetic resonances produced at $\Omega_m$ =60 kHz ($\Omega_L$) and $\Omega_m$=180 kHz ($3\Omega_L$) due to Λ-systems formed by strong ($\sigma^\pm$, π) excitations [shown in Fig. 2(c)]. Formation of $3\Omega_L$ resonance is unique to this approach. It can be used to exactly establish the two-photon resonance condition (i.e. Δ=0) between the two laser fields. Direction of the external magnetic field can also be determined by using the amplitude ratio between $\Omega_L$ and $3\Omega_L$ resonances, and pre-calibrating the variation of this ratio with light ellipticity and polarization angle [21]. The transverse field $B_y$ is expected to produce a very weak resonance at $\Omega_m$=120 kHz ($2\Omega_L$) due to Λ-systems formed by weak ($\sigma^+$, $\sigma^+$) or ($\sigma^-$, $\sigma^-$) excitations [shown in Figs. 2(a), 2(b)]. This resonance is not visible in the experimental Fig. 4a (blue curve). To further understand the origin of magnetic resonances produced by the synchronous CPT scheme, we developed a theoretical model based on the density matrix equations by considering only three hyperfine states (i.e. $F_g$=1, $F_g$=2 and $F_e$=2) in Na $D_1$ line [22]. We further simplified our model by neglecting the effect of atomic motion, and the spatial distribution of light intensity. Figure 4b (blue curve) shows calculated magnetic resonances at $\Omega_m$ = $\Omega_L$ and $3\Omega_L$. Unlike the experiment [Fig. 4a (blue curve)], a weak magnetic resonance at $\Omega_m$ = $2\Omega_L$ is observed using the theoretical model.

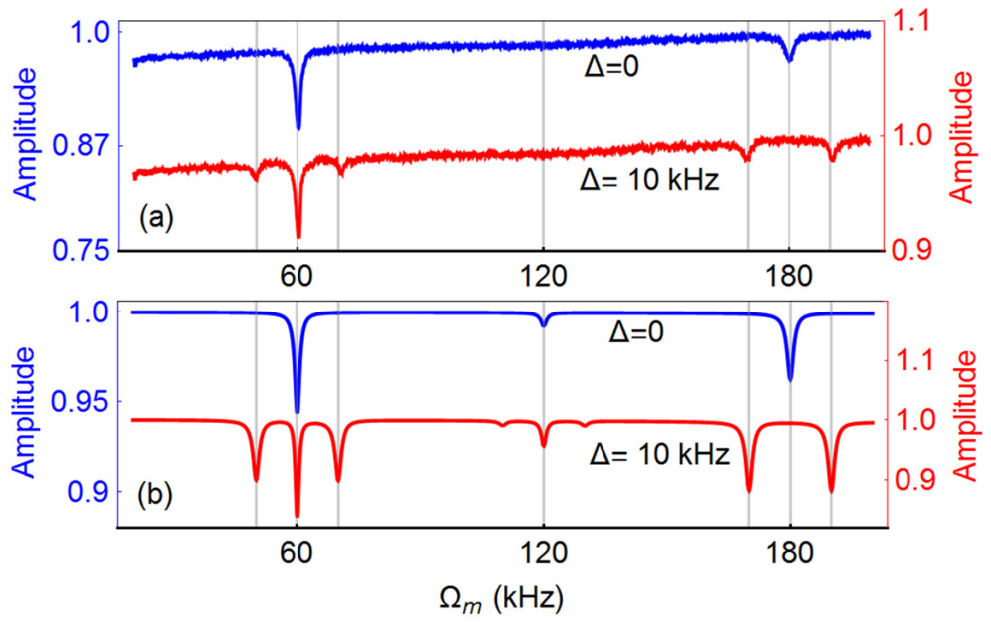

Fig. 4. (a) Experimentally and (b) theoretically obtained magnetic resonances as a function of $\Omega_m$ for two different Δ values. Parameters used in the simulations: Rabi frequencies $\Omega_c$=$2\Omega_s$ = 0.03 Γ and transit decay rate τ =$10^{-5}$ Γ, where Γ is the spontaneous decay rate.

When a difference detuning Δ=10 kHz is introduced between the light fields, two-photon resonance conditions are also satisfied for $\Omega_m$=$\Omega_L$±Δ and $\Omega_m$=$3\Omega_L$±Δ, thus, forming new resonances around $\Omega_m$=$\Omega_L$ and $\Omega_m$=$3\Omega_L$ resonances (Figs. 4a, 4b: red curves). In this case, a two-photon resonance condition cannot be satisfied at $\Omega_m$ =$3\Omega_L$ by the two light fields with Δ=10 kHz. Hence, no $\Omega_m$ =$3\Omega_L$ resonance is observed in Fig. 4(a,b) for Δ=10 kHz case. However, $\Omega_L$ resonance can still be seen for Δ=10 kHz. This can be explained by the formation of $\Omega_L$ resonance due to single-photon resonance in the Bell-Bloom scheme [9] through $\Lambda_{12}$-system, shown in Fig. 2(d). Our theoretical calculations (Fig. 4b) show similar results as in the experiment (Fig. 4a). The theoretical plot shows a weak resonances at $\Omega_m$= 120 kHz ($2\Omega_L$) formed through $\Lambda_{11}$-system, shown in Fig. 2d for Δ=10 kHz case.

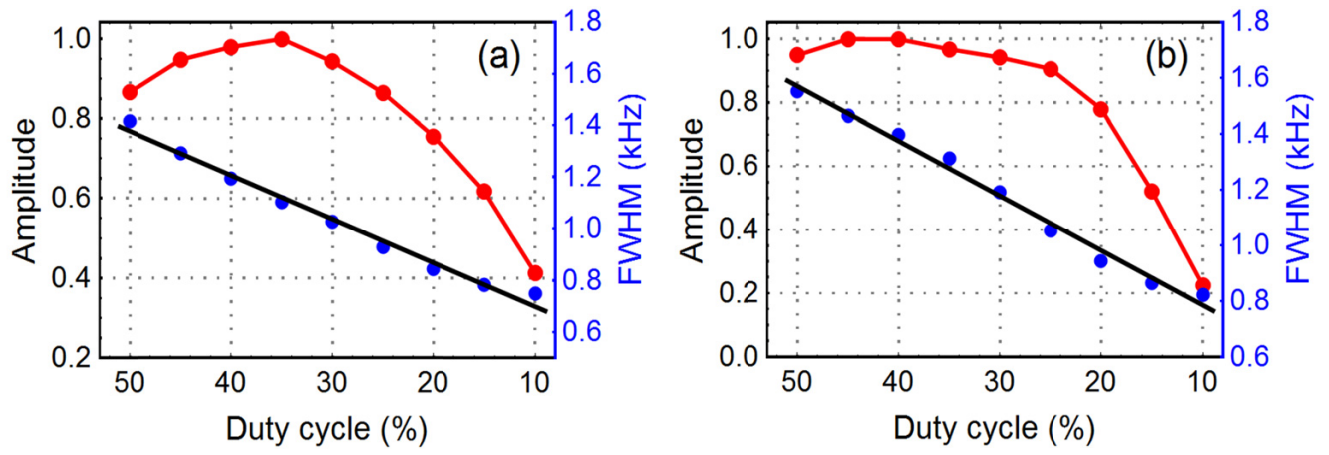

Fig. 5. (a) Experimentally and (b) theoretically measured peak amplitude and FWHM of $\Omega_m$=$\Omega_L$ (180 kHz) resonance. Black lines in both plots show linear fittings to the linewidth data. Other simulations parameters are same as in Fig. 4(b).

Finally, we demonstrate an atomic magnetometer for remote magnetometry application, based on the $\Omega_L$ resonance signal produced by the synchronous CPT scheme. Figure 5(a) shows experimentally measured peak amplitudes (red dots) and widths (blue dots) of $\Omega_L$ resonance at different duty cycles of CPT pulses. The peak intensity in CPT fields is kept constant at 5.2 W/m$^2$ for all measurements. A transverse field $B_y$ = 257 mG, comparable to earth field, is applied and the modulation frequency $\Omega_m$ is scanned around the Larmor frequency $\Omega_L$=180 kHz. The resonance signal is acquired using an LPF with cut-off frequency 1 kHz. As the duty cycle is lowered from 50%, the amplitude of $\Omega_L$ resonance initially increases, and maximizes near 35% duty cycle (Fig. 5a: red curve). Thereafter, the amplitude deceases almost linearly over the duty cycle ranging from 30% to 10%. Unlike $\Omega_L$ resonance produced in the Bell-Bloom scheme [8], synchronous CPT scheme delivers high amplitude $\Omega_L$ resonance at longer duty cycles closer to 50% [Fig. 5(a)]. Long duty cycle corresponds to long CPT pulses, which are beneficial in sky experiments for increasing the return fluorescence from mesospheric sodium layer. To keep the peak intensity constant at 5.2 W/m$^2$, the average intensity in CPT pulses is lowered proportionately with the lowering of the duty cycle. Therefore, the linewidth (i.e. full-width at half-maximum, FWHM) of $\Omega_L$ resonance reduces linearly with duty cycle [Fig. 5a: black curve]. Theoretical results shown in Figure 5(b) qualitatively agree with experimental results shown in Figure 5(a).

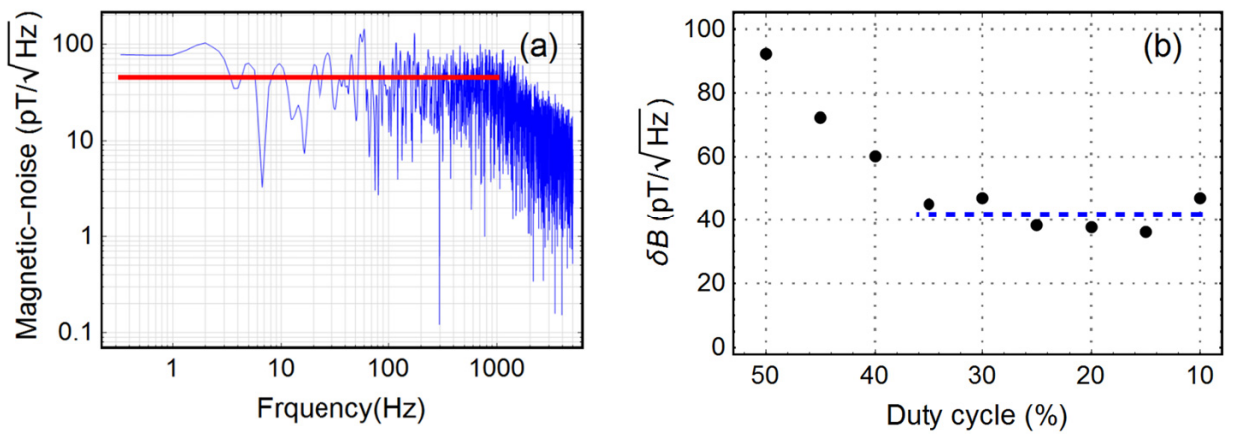

Fig. 6. (a) Magnetic field noise spectrum in $\Omega_L$ resonance signal, measured at 35% duty cycle. The red line shows the noise-floor in the frequency range 1Hz to 1 kHz. (b) Variation in sensitivity $\delta B$ as a function of the CPT pulse duty cycle. The dashed blue line shows the average sensitivity from 35% to 10% duty cycle.

The sensitivity of a magnetometer depends on the amplitude and the linewidth of resonance. The photon shot-noise limited sensitivity $\delta B$ of the magnetometer is given by

$$\delta B = \Delta f / (\gamma . S/N) \quad (1)$$

where $\Delta f$ is the FWHM of $\Omega_L$ resonance, $\gamma$ = 6.99812 Hz/nT is the gyromagnetic ratio of sodium atom, and $S/N$ is measured in $\sqrt{Hz}$ by measuring the noise in the $\Omega_L$ resonance signal using a fast-Fourier-transform (FFT) on the oscilloscope. Figure 6(a) shows the magnetic noise spectrum using 35% duty cycle in modulation. The root-mean-square of magnetic noise is measured around 45 pT/$\sqrt{Hz}$ within the frequency range 1 Hz -1 kHz (Fig. 6a: red line). Since the LPF is set to a 1 kHz cut-off frequency, the noise decreases rapidly for frequencies above 1 kHz. Figure 6(b) shows the variation in sensitivity $\delta B$ with duty cycles. The sensitivity $\delta B$ improves nearly by a factor of two, as the duty cycle is decreased from 50% to 35%, where amplitude of resonance is maximum. For lower duty cycles below 35%, the sensitivity remains constant without much variation. An average sensitivity of 41 pT/$\sqrt{\mathrm{Hz}}$ (Fig. 6b: dashed line) is achieved for duty cycles ranging between 35% and 10%. A maximum sensitivity close to 37 pT/$\sqrt{\mathrm{Hz}}$ is achieved for 25% duty cycle. In the reported literature, highest sensitivity in sky experiment is recorded at 20% duty cycle, employing the Bell-Bloom scheme [11]. The synchronous CPT scheme offers higher sensitivity at much longer duty cycles. Additionally, the dark $\Omega_L$ resonance in sodium $D_1$ line is associated with reduced photon shot noise, and therefore, expected to give better magnetic field detection sensitivity in sky experiments [7]. Besides, the sensitivity $\delta B$ measured in the current setup is sub-optimal. It can be further improved by simple modification of the modulation technique. The amplitude of $\Omega_L$ resonance can be maximized by equalizing the power in the CPT fields. This can be achieved by configuring the EOM differently such that the $\Omega_L$ resonance is produced by the two first-order sidebands in the presence of a highly suppressed carrier. Our experiment is also done in a weakly shielded environment, and therefore, the measured sensitivity $\delta B$ may have been limited by the fluctuations in the ambient magnetic field. We expect that improving the shielding environment will further improve the sensitivity of the magnetometer to sub-picotesla level, closer to the photon shot-noise limited sensitivity.

In summary, we have developed a new synchronous CPT scheme, showing the possibility for improving the sensitivity in remote earth field measurement. This technique uses two-photon resonance instead of single-photon resonance used in the Bell-Bloom scheme for remote magnetometry. We showed that, with two resonant fields, atoms in both ground states contribute to produce a strong $\Omega_L$ resonance. The proposed technique significantly enhances the sensitivity in magnetic field measurement at longer duty cycles than previously reported. An average magnetic field sensitivity of 41 pT/$\sqrt{\mathrm{Hz}}$ is measured with pulse duty cycle ranging from 35% to 10%. We expect to conduct remote magnetometry experiments employing this technique in sodium $D_1$ line in the near future.

**Acknowledgement:** This work was supported by NASA EPSCoR award #80NSSC17M0026, and NASA MIRO award #NNX15AP84A.

**Disclosures.** The authors declare no conflicts of interest.